# Nanoscale force sensor and actuator with carbon-nanotube network

Zhiping Zhang[1], Li Huang[1], Haifei Wang[1,2], Tracy Bay[2], Robinson L. Smith[2]

[1]Changzhou University, Changzhou, China, [2]Rio Salado College, Tempe, AZ, USA

## ABSTRACT

*Actuators at the nanoscale are assuming great important in today's shrinking electronics industry where a system-on-chip approach is used to integrate multi-functional systems and devices. Integration of sensors, actuators, information processing and storage have become extremely important, more so at the sub-micrometer scale. Here we utilized vertically aligned carbon nanotube forests bonded onto a piezoelectric substrate to provide sensing of nanoscale forces and actuation on the same platform. This is a demonstration of integration of two important functions using the functionality of nanomaterials. This has immediate utility in many areas including outer space exploration, portable consumer electronics, embedded systems, etc.*

**Keywords:** Carbon nanotubes, Frequency Modulation (FM), Pressure Sensing, Local actuation, Low Pressure, Capacitive Transduction, Sensing Membrane, Surface Micro Machining,

## 1. INTRODUCTION

Structures with multi-functional capabilities including actuation, sensing, diagnostics, self-mitigation, etc. are highly sought after for critical applications in a variety of areas including structural safety, space exploration, embedded electronics, etc. Sensing and actuation are two very important functions which need to be addressed in such systems. Here we demonstrate pressure sensing and force actuation in a nanoscale system by utilizing carbon nanotubes, one of the most popular nanoscale systems.

Piezo electric pressure sensor, mechanical deflection of a spring system, capillary barometers, determining membrane deflection using optical methods, bourdon tubes or bellows, piezo resistive pressure sensor, capacitive pressure sensor are some of the popular methods of sensing pressure.

Attenuation of output signal, since amplitude of output is of prime importance, is a major disadvantage especially in sensing low pressures. There is a need for more stages of voltage amplifiers, noise filtering mechanisms, buffers for impedance matching, etc in the methods described above. Noise introduced in the circuit affects the output shape of signal (waveform). Distortions introduced internally in the circuit and due to temperature variations affect accuracy of results. These issues lead to more stages of noise filtering mechanisms. More stages lead to more parasitic components, complicating the exact analysis of the circuit. Transmitting mechanism would be separately needed when output has to be sent to a receiver as a modulated signal. Dynamic pressure cannot be applied above a certain frequency because of limitations in the sensing membrane. Also to be noted are the serious sensitivity issues in design when designed for low pressures leading to sophistication and high cost. Generally found in the above mechanisms are highly pronounced shortcomings in frequency management, cropups through noise, crosstalk, signal dropout, complete transmission interruption affect the transmission of output.

It is also of particular interest to have a reliable system of transduction in order to actuate the surface of a desired system with accurately known stresses and displacements. This is useful in calibrating the surface stress measurement system and in also determining the limits of the stresses that the given material can withstand. When the surface stress measurement system is to be negligibly small and also negligibly heavy compared to the system of which it senses the surface stresses, there is a need to find materials that suit this requirements, unlike using integrated sensing membranes and other related sensing structures.

## 2. DESIGN OF THE STRUCTURE

A membrane area of 2.6 µm X 2.6 µm was chosen. This is a tradeoff. This was carefully chosen to keep the cavity capacitance higher than any parasitic capacitances (few pF), to keep the membrane as small as possible (to take advantage of miniaturization and to see that maximum pressure felt by it does not exceed burst pressure) and also to take care of the resonant frequency of the membrane, which is to be about 3 to

4 times the input frequency. Also note that a square membrane was chosen over a rectangular design. A rectangular membrane, of maybe a high aspect ratio, would have produced a better deflection for the same pressure applied, as in [1], but the change in the capacitance of the device would still remain the same. Hence for ease in design and analysis, a square membrane was chosen.

The gap between the capacitor plates was taken to be 0.5 to 1 µm. (air gap).The membrane thickness was chosen to be 0.5 to 1µm. (made of poly silicon). Upon this, a piezoelectric layer was deposited followed by growth of carbon nanotubes, as described in our earlier investigations.

The advantage of using a material like a piezo electric material is that the output is obtained without biasing the system with any activating energy of signal. Stress is a measure of the average amount of force exerted per unit area of the surface on which internal forces act within a deformable body. It is a measure of the intensity, or internal distribution of the total internal forces acting within a deformable body across imaginary surfaces. These internal forces are produced between the particles in the body as a reaction to external forces applied on the body. External forces are either surface forces or body forces. In general, however, the stress is not uniformly distributed over a cross section of a material body, and consequently the stress at a point on a given area is different than the average stress over the entire area. Therefore, it is necessary to define the stress not at a given area but at a specific point in the body.

Piezoelectricity is the ability of some materials (notably crystals and certain ceramics, including bone) to generate an electric potential in response to applied mechanical stress. This may take the form of a separation of electric charge across the crystal lattice. If the material is not short-circuited, the applied charge induces a voltage across the material. The piezoelectric effect is reversible in that materials exhibiting the direct piezoelectric effect (the production of electricity when stress is applied) also exhibit the reverse piezoelectric effect (the production of stress and/or strain when an electric field is applied). For example, lead zirconate titanate crystals will exhibit a maximum shape change of about 0.1% of the original dimension.

## 3. TESTING AND RESULTS

With this technique, frequency changes can be made appreciable and sensed easily even for low pressures unlike small changes in output voltage or current. Low electrical self-noise and noise cancellation are possible by using an array of micro devices in a small region and micro sized devices will have lower noise induction. Micro size with silicon fabrication allows easy integration with microelectronics. Micro size also allows for applications like surveillance, multiple transducers in a small area, in ear translators and also makes good mechanical filters. With a good design of the membrane, high sensitivity of desired magnitude can be achieved. Good fidelity of the signal is achieved in relative ease compared to other technologies. Low impedance, low power consumption and hence temperature independence can be achieved to an appreciable extent. The carbon nanotube forest can be designed for wide frequency response. And in this case the input frequency can be considerably high.

Incorporating frequency changes for sensing pressure (which is not an industrial norm, unlike the piezoresistive and capacitive sensing, which are quite the norm) is stressed in this paper. A variable capacitor is used to sense input pressure. And when this is incorporated in the tuning circuit of an oscillator, it gives rise to changes in output frequency. (frequency being the output has significant advantages over other methods of measurements). Proposing the use of a micro device for the tunable capacitor, foreseeing the various advantages involved in this technology and also the advantages over the existing MEMS technologies for sensing mechanisms are discussed here.

The device was next tested with an available tunable capacitor (0.28 μF). This capacitance was separately measured to vary from 0.28 to 0.35 μF, and by slight blowing to impinge the applied pressure on the membrane. The applied pressure was acoustic, by blowing air slightly. This pressure was calculated using an electronic balance and by blowing on the same area on it and then reading the change in weight measured (this area was created by blowing into a hole made in a bulk of sponge placed on the balance). This was found to be approximately 50 Pa for a mild breath (L=10 μH, C= the variable capacitor= 0.28μF). The circuit is a standard "Hartley's oscillator", using a bipolar junction transistor (BJT) circuit which needs no explanation.

These capacitors were similar to those tunable capacitors which cannot be used to generate oscillations in a tank circuit. So these had to be carefully coupled with ceramic capacitors of lower values in parallel with them. (Like 0.28 μF with 0.001 μF). Since this is just an LC circuit, with a BJT, there was a limit on the maximum frequency achieved (about 1.11 MHz). The use of BJT also induces distortions in the wave shape in the output when operated at around 400 kHz. This explains the slightly distorted waveform in the oscilloscope output. Since the mechanism of these prefabricated capacitors is designed for measuring amplified changes in output voltage, the sensitivity of the capacitor to input pressure was lower than desired. The high magnitude of capacitances (μF) also restricts the frequency range.

This approach tackles almost all the above issues discussed above by the virtue of usage of frequency modulation and miniaturization. There is to a great extent, retention of fidelity of the modulating wave even after the modulated wave attenuates. This also avoids the need for many amplifier stages. Noise will not affect frequency sensing to any significant magnitude. Distortions induced in the wave shape will not affect the frequency readout (as this is FM). This also avoids the use of many amplifier stages since output is only frequency readout.  Since output is FM, transmission, if necessary, can be done simply by using a power amplifier and an antenna after this single stage circuit.

We remark the high signal intensity observed upon pressure actuation. The signal to noise ratio (SNR) was about 2000, which is one order of magnitude better than previous results using similar techniques. We found that as the pressure increased, the non-linear output from the piezoelectric surface in contact with the carbon nanotubes provided a negative voltage response. This we consider to have originated from the reverse piezo effect. We found that bonding the CNTs to the piezo substrate at a temperature of 1000 K was the most suitable condition for the best transfer of the CNT layer. Also, among the different modes represented in Figure 5, we find that the A0 signal is usually higher than the S0 signal.

## 4. CONCLUSIONS

The device described here is suitable for sensing low pressures, for example, the pressure developed in the brain after head injury. (about a 100 Pa) (the devices used to measure this low pressure cost quite a lot, whereas this device costs very less). The slight variations in pressure in the atmosphere that have to be sensed accurately by the meteorology labs need such sensitive devices. There is a need for constant pressure monitoring of fluid pressure in fuel manufacturing plants to see that there is no danger due to leakage of fuels. This employs sensitive pressure sensors. This device discussed will be ideal for constant monitoring of fluid pressure also. Aeronautics is another domain which uses sensors and controllers extensively. This low pressure sensor will be very useful to sense slight variations in atmospheric pressure outside the aero plane to detect the occurrence of turbulence if any. (i.e.: this can replace the existing ones because this is more sensitive and also cost effective). The micro size also allows for applications like surveillance, in-ear translators, etc.

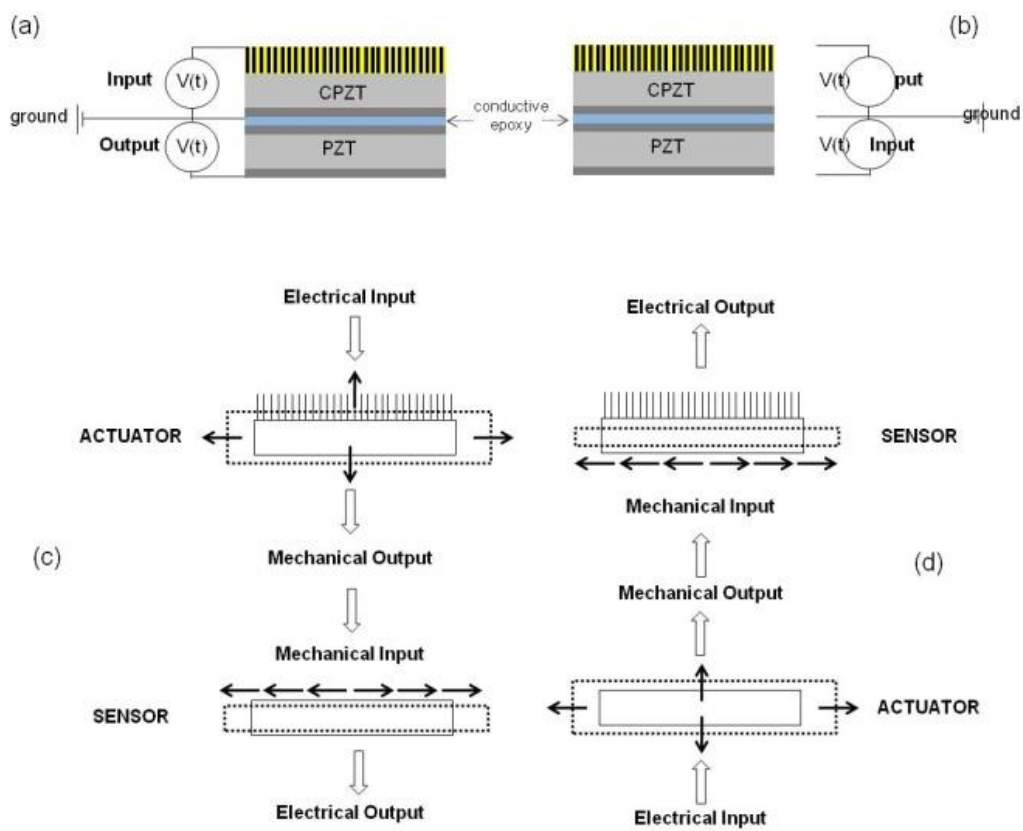

Figure 1: (a), (b) Schematic of the piezoelectric-carbon nanotube stack used for sensing and actuation. (c), (d) Schematic illustration of the physical origins of the working.

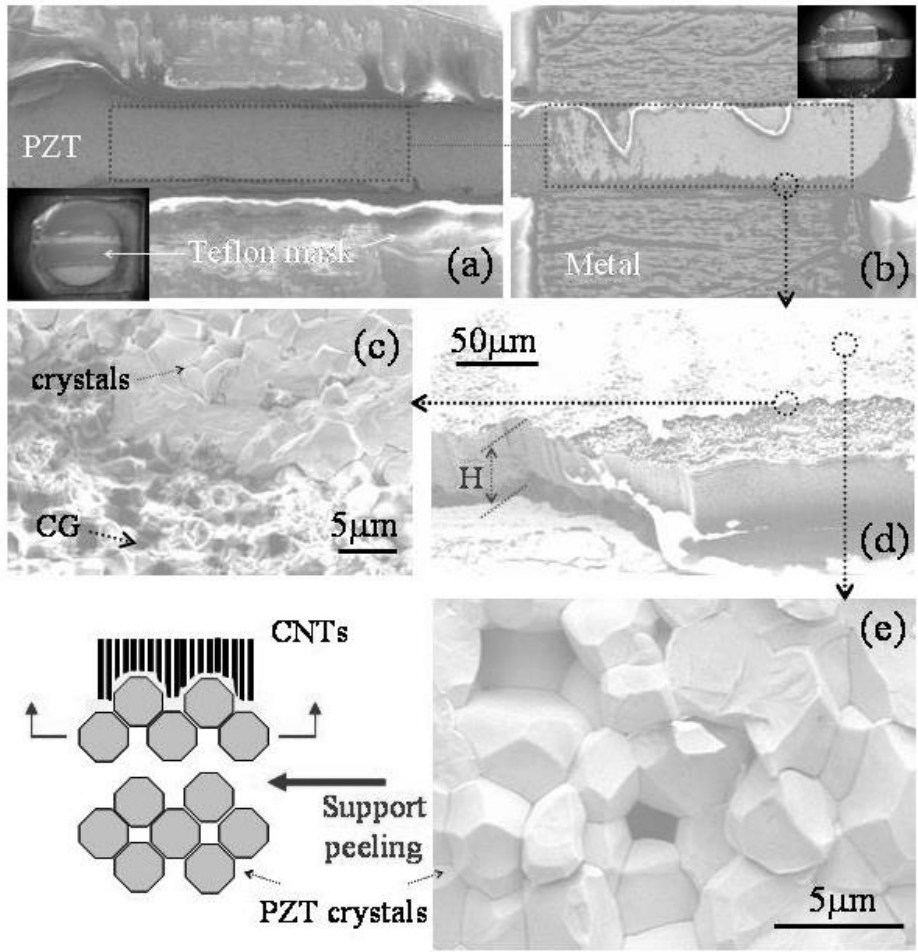

Figure 2: (a) SEM of the piezoelectric surface, (b) SEM of the metal surface, (c) top view of the interface with the piezoelectric (d) Image of the nanocomposite bond on the metal surface (e) magnified piezoelectric crystals peeled from the piezoelectric substrate.

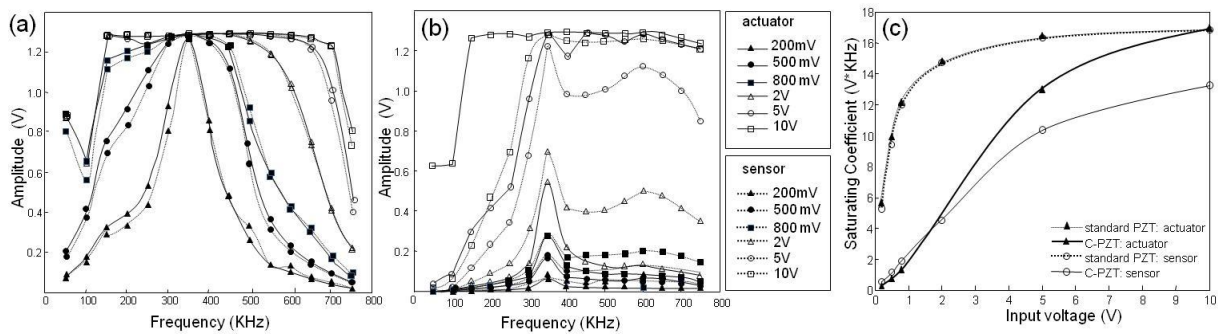

Figure 3: Stack piezoelectric test results. (a) Standard piezoelectric; (b) piezoelectric with CNTs electrode; (c) Saturating coefficient and voltage for the actuating and sensing case of a piezoelectric with and without CNTs.

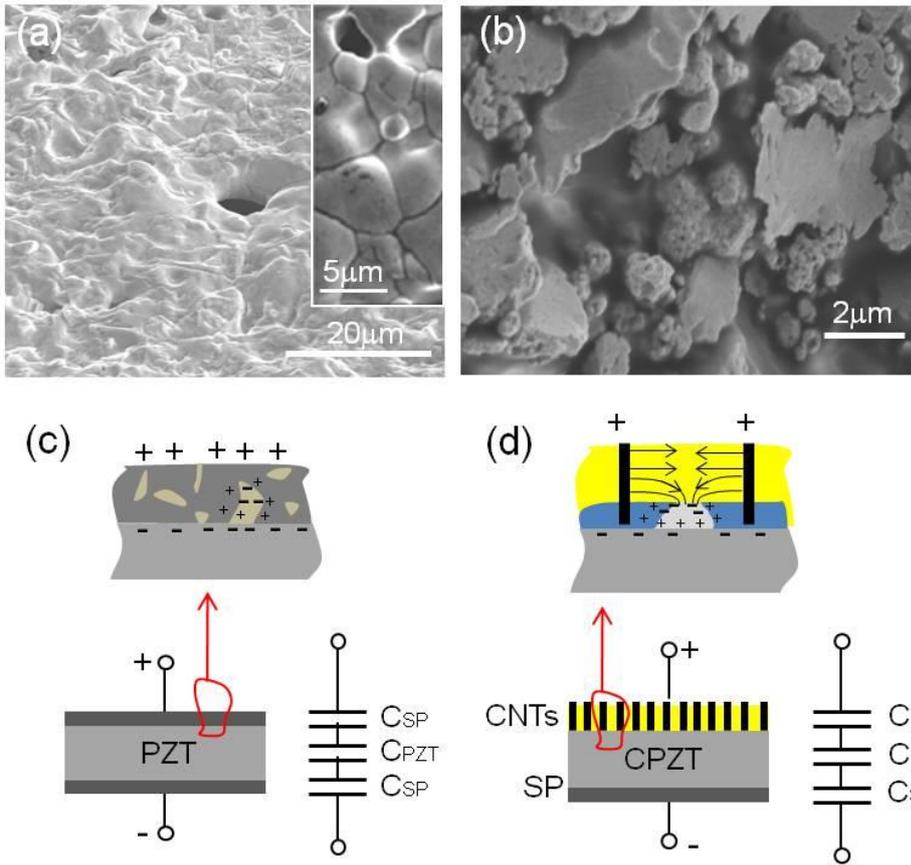

Figure 4: (a) SEM of the piezoelectric surface with no carbon nanotubes and (b) SEM of the carbon nanotube surface with piezoelectrics. (c) Description of the parasitic capacitances in the structure and (d) capacitances within the CNT layer.

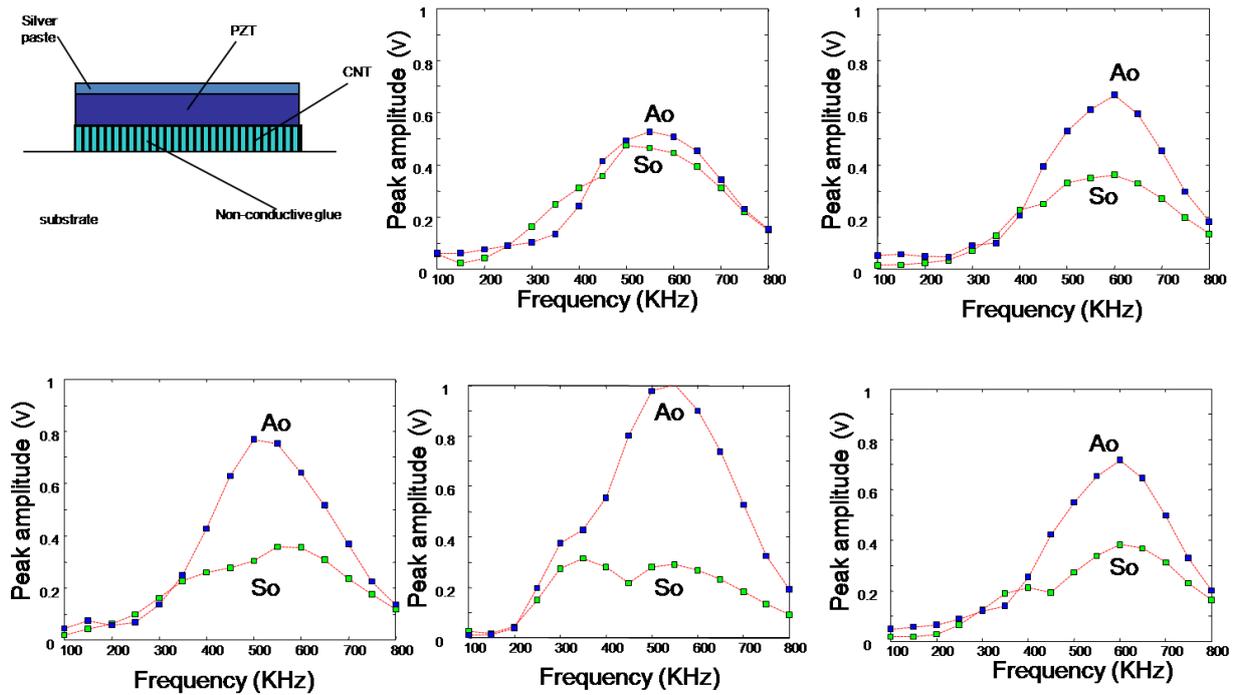

Figure 5: Output signals for different modes (A0 and S0) from the piezoelectric materials with and without CNTs.

## Rreferences